# The connection between vortex-like topological excitations and conventional excitations in quantum ferromagnetic spin systems on two dimensional lattice and their stability

Subhajit Sarkar
subhajit@bose.res.in

Ranjan Chaudhury
ranjan@bose.res.in

Samir K. Paul
smr@bose.res.in

S. N. Bose National Centre For Basic Sciences,
Block- JD, Sector- III, Salt Lake, Kolkata 700098, India.

**Abstract:** We present a scheme for the construction of quantum states of vortex like topological excitations corresponding to spin- ½ strongly XY anisotropic nearest neighbor Heisenberg Ferromagnet on two dimensional lattice. The procedure involving Pauli spin basis states is carried out corresponding to both infinite dilute limit and finite density limit of vortex/anti-vortex. It is found that the corresponding quantum mechanical states representing charge 1 quantum vortices/ anti-vortices can be expressed as linear combinations of single magnon states, composite multi-magnon states and the ground state. Detailed calculations show that these states are quantum mechanically stable states of the Hamiltonian only when the system size exceeds certain threshold value. Our analysis indicates that the interactions between different magnon modes can very well generate these topological excitations. Possible applications of our calculations to real magnetic systems are also discussed. Magnetic measurements probing spin dynamics may be undertaken to verify the existence of the threshold size for the stability of vortices.

# 1 Introduction

In magnetic systems in low dimensions, viz., one dimension (or rather quasi-one dimension) and two dimensions (or rather quasi-two dimensions), the occurrences of topological excitations of solitons and vortices/merons respectively, are natural as they are thermodynamically feasible [1-11]. In these systems vortices/merons play a crucial role in bringing out a non-conventional phase transition in the two spatial dimensions [12].

In the last few years there has been a renewed research interest especially in the quasi-two dimensional magnetic systems motivated by the aims of building magnetic devices. These devices make use of mobile vortices [13-15]. In the magnetic thin films the interplay between the exchange interaction and the magnetic dipole - dipole interaction causes the formation of domain structures in absence of magnetic fields. Furthermore, each of these domains contains a magnetic vortex characterized by curling in-plane magnetization located at the center. The component of magnetization perpendicular to the plane of the film serves as 'Polarization' of the vortex core [16]. Such a magnetic vortex has been proved to be a potential candidate for switching devices as well as for data storage where the 'Polarization' of the core can be manipulated in a controlled manner [16]. Direct experimental evidences of such vortex states have been verified by Magnetic Force Microscopy (MFM) and also by the spin-polarized Scanning Tunneling Microscopy (STM) [17, 18].

The concept of topological phase transition driven by binding to unbinding processes involving vortex, anti-vortex like topological excitations in two dimensional ferromagnetic systems was introduced by Kosterlitz and Thouless and independently by Berezinskii [19, 20]. Their ideas backed up by analytical and numerical studies led to the signature of the above transition (BKT) in the dynamical correlation function, in two dimensional Heisenberg ferromagnetic models of both XY-type and XY-anisotropic type [21-26]. Attempts were also made later to extend the proposal of BKT to case of XY-anisotropic anti-ferromagnetic two dimensional systems [24, 27, 28]. The phase transition described above occurs at a temperature $T_{BKT}$ characterized by degree of anisotropy. Above this the dynamics of freely moving vortices and anti-vortices (or merons and anti-merons) provides non trivial contribution in the dynamical correlation function as mentioned above, giving rise to the well-known "central peak". In the Inelastic Neutron Scattering experiment (INS) the existence of such a central peak at $\omega = 0$ has been observed in the plot for the dynamical structure function $S(\boldsymbol{q}, \omega)$ vs. neutron energy transfer $\hbar\omega$ in the constant 'q' scan [8, 9]. The materials on which such experiments have been performed include different layered ferromagnetic and anti-ferromagnetic materials such as $K_2CuF_4$, $La_2CuO_4$, $CuGeO_3, BaNi_2V_2O_8, BaNi_2(PO_4)_2$ and stage-2 $CoCl_2$ graphite intercalation compound [8-11, 29-34]. The existence of BKT transition in these materials has been proposed on the basis of investigations performed on these materials using both INS and the Electron Spin Resonance (ESR) techniques.

Some attempts have been made in the past to describe the dynamics of mobile vortices or merons corresponding to XY or XY-anisotropic Heisenberg model assuming that these topological excitations constitute a classical ideal gas [21-24]. Approximate analytical theories and Monte Carlo molecular dynamics (MCMD) simulations have suggested that the existence of the central peak in the dynamical structure function is partly due to the scattering of neutrons from the above mobile topological excitations [21-24]. However, the semi-classical ideal vortex gas phenomenology turns out to be quite inadequate even for ferromagnets, in explaining the experimental results corresponding to systems having low spin values, even after incorporation of suitable quantum corrections in the calculations [35]. Although the occurrence of central peak is ensured, the theoretical dynamical structure function turns out to be negative for a considerable range of $\omega$, when the appropriate experimental conditions are taken into account in the calculations! This very peculiar feature clearly signals total failure of the semi-classical treatment based on ideal gas phenomenology corresponding to low spin systems. Therefore a full-fledged quantum mechanical description and treatment of such topological spin excitations become very crucial [35]. Besides, none of these semi-classical approaches so far have incorporated any spin wave-vortices/merons interaction and rather assumes that the vortex/merons take the shapes of spin profiles independent of spin waves.

The question of existence of the topological excitations, namely, vortices and merons in two-dimensional quantum ferromagnetic spin systems have been explored both numerically and analytically [36-40]. It has been determined numerically that in this case the vortex- anti-vortex pair density is nonzero even at T=0 [36]. In a pure quantum mechanical treatment it has been found that almost all the vortices and anti-vortices are bound in pairs on square lattice and the number of isolated free vortices per site vanishes for $T < T_{BKT}$ [36]. Monte Carlo simulations have also been performed on quantum XY model on two-dimensional lattices. The validity of the BKT transition for this model has been confirmed [37, 38]. A full-fledged quantum treatment has also been performed based on the application of path integral techniques using the coherent state basis, for XY- anisotropic Heisenberg ferromagnet on a square lattice [39, 40]. The partition function for the above quantum spin model has been expressed in terms of an effective action containing a topological part (Wess-Zumino term) which contains a genuine topological term as a charge measuring object for the vortices/merons (anti-vortices/anti-merons) alongside a non-topological term. It has been shown that in the very large anisotropy limit (corresponding to $\lambda \to 0$) the topological term can characterize the topological excitations viz., vortices and anti-vortices [39, 40]. In this formalism, the topological term arises from the path integral formulation of the quantum partition function in contrast to the situation where the vorticity operator has been introduced heuristically [36].

Incidentally, a rather different approach has been put forward to explain the origin of the central peak in the dynamical structure function $S(\boldsymbol{q},\omega)$ corresponding to XY anisotropic classical Heisenberg ferromagnetic model in two spatial dimensions. In this theory, the occurrence of the peak has been attributed to the fluctuations of the density of the topological excitations due to local diffusion and creation-annihilation of merons and anti-merons [41].

The behavior of the collective modes like spin waves in the presence of a single vortex/meron corresponding to two-dimensional easy plane classical Heisenberg ferromagnet have been investigated using approximate analytical treatment in the continuum limit and numerical diagonalisation techniques. It is found that the renormalized spin wave modes show a strong localization of their amplitudes near the vortex core [42-44].

In this article we describe our investigations on the possible composition of these topological excitations of true quantum nature in low-dimensional anisotropic quantum Heisenberg ferromagnetic model. It turns out that the interactions between the different multi-magnon modes play a very important role in the formation of the above excitations. These multi-magnon interactions are generally neglected in the linear spin wave/ one-magnon theory and even in the BKT theory. Theoretical attempts were made afterwards to study the interplay between classical spin waves and vortices (merons) in the regime $T < T_{BKT}$ [25]. These treatments lead to a renormalization of the exchange coupling without any explicit spin-wave - vortex coupling [26]. Magnon modes are low energy excitations and represent a quantized coherent precessional motion of all the spins around the direction of the spontaneous magnetization in the long range ordered phase. These modes however, become ill-defined in the short range ordered phase [45-48]. In contrast, the quantum states representing topological spin excitations are found to be stable even in the short range ordered phase when the system size is very large, as we will demonstrate in this communication.

The article is organized as follows:- In section 2, we explain our mathematical formulation by constructing the quantum state corresponding to 1-vortex (and 1-anti-vortex as well) in the strong anisotropy limit of the XXZ model and establish the connection between the quantized vortex states and the magnon states; in section 3, we analyze the quantum mechanical stability of such vortex/anti-vortex states for both the cases of infinite dilute limit and the finite density limit; finally in Section 4, we present the conclusions and discussions and also highlight the possible application of the results of our present investigation to the real magnetic systems.

## 2 Mathematical formulations

Most of the material systems showing the so called "central peak" (described in the introduction), are governed by the XY anisotropic quantum Heisenberg (XXZ) Hamiltonian,

$$\mathcal{H} = -J \sum_{\langle ij\ pq \rangle} \left( S_{ij}^x S_{pq}^x + S_{ij}^y S_{pq}^y + \lambda S_{ij}^z S_{pq}^z \right) \tag{1}$$

on a two-dimensional square lattice with nearest neighbor interaction, where $\lambda$ ($0 \leq \lambda < 1$) is the anisotropy parameter and for ferromagnetic systems $J > 0$. Here $S_{ij}^x, S_{ij}^y$ and $S_{ij}^z$ are the $x, y$ and $z$ components respectively of the spin operator on the $ij$-th lattice site. We shall concentrate on the $S = \frac{1}{2}$ ferromagnetic model in the very strongly XY- anisotropic limit ($\lambda \to 0$, i.e., $\lambda$ is vanishingly small, but $\lambda \neq 0$). With this smallest '$S$' value, the model is infact in the extreme quantum regime.

It is worthwhile to mention here that the classical counterpart of the above model admits of the well-known meron solution [21-24, 49, 50]. In classical case spins are considered to be classical vectors of magnitude '$S$' in spherical polar coordinate. The static meron solution corresponding to the classical version of model (1) is given by,

$$\phi(x,y) = \pm \arctan\frac{y}{x}, \qquad \theta(x,y) = \frac{\pi}{2}\left(1 \pm e^{-\frac{r}{r_v}}\right) \text{for } r \gg r_v, \tag{2}$$

$$= 0 \text{ or } \pi \text{ for } r \to 0,$$

for single meron centered at $r = (0,0)$, where $\phi(x,y)$ is the azimuthal angle and $\theta(x,y)$ is the polar one and $r_v$ is the meron core radius [21-24, 51]. The eqn. (2) describes the asymptotic behavior $\theta(x,y)$. Numerical studies have led to the conclusion that there is a critical value of the anisotropy parameter, say $\lambda_c$, below which only the static flattened merons or ordinary vortices are stable and above that the normal merons are stable [49, 50].

Let us now come back to the properties of the quantum ferromagnetic XY anisotropic Heisenberg model (ferromagnetic XXZ model) at very low temperature. First of all this model is well known to possess its ground state (and eigen-state as well) having spins all aligned along the '+ve' or '-ve -z' direction The normalized ground state $|0\rangle$ for the Hamiltonian in the case of ferromagnetic model as in eqn. (1) is chosen along the negative Z axis and is defined as $S_{ij}^- |0\rangle = 0$ for every $i,j$. Explicit form for the ground state on the square lattice is given by,

$$|0\rangle = |\downarrow\rangle_{11} \otimes |\downarrow\rangle_{12} \otimes |\downarrow\rangle_{13} \otimes \cdots \otimes |\downarrow\rangle_{ij} \otimes |\downarrow\rangle_{i+1\,j} \otimes |\downarrow\rangle_{i+1\,j+1} \otimes |\downarrow\rangle_{i\,j+1} \otimes \cdots \otimes |\downarrow\rangle_{N\,N} \tag{3}$$

for a $N \times N$ square lattice, where $S_{ij}^-$ is the spin lowering operator defined as $S_{ij}^- = S_{ij}^x - iS_{ij}^y$. The lattice has the structure of a torus for periodic boundary conditions [51-56]. The ground state energy is denoted by,

$$\mathcal{E}_0 = -\frac{\mathcal{N}}{2}\lambda J\hbar^2, \tag{4}$$

corresponding to the ground state $|0\rangle$, where $\mathcal{N} = N^2$.

It is important to recall that a quantum Heisenberg model (ferromagnetic or anti-ferromagnetic) on a three dimensional lattice exhibits long range ordering at finite temperature unlike its counterparts in one and two dimensions[1] [57-59]. In the three dimensional case the collective excitations, viz., magnons are well defined in the long ranged ordered phase and become fragile in the short ranged ordered phase above the transition temperature, Curie temperature ($T_c$) for ferromagnetic systems or Néel temperature ($T_N$) for anti-ferromagnetic systems [45-48]. In analogy with the above three dimensional case, it is expected that in two-dimensions we can still find some fragile magnon-like excitations along with multi-magnon composites within a very small temperature regime above zero [45-48].

---

[1] According to Mermin-Wagner (MW) theorem (see [57-59]), at any nonzero temperature, a one- or two-dimensional isotropic spin-'S' Heisenberg model with finite-range exchange interaction cannot exhibit any long range ferromagnetic order (implying $T_c = 0$) or antiferromagnetic order (implying $T_N = 0$). It can be shown that even for XY anisotropic Heisenberg Ferromagnet/Anti-ferromagnet on two-dimensional lattice the deviation from the spontaneous magnetization has infrared logarithmic divergence. This implies that the MW theorem holds for two dimensional XY anisotropic Heisenberg Ferromagnet/Anti-ferromagnet as well.

However, as a first approximation we consider, in our present analysis for the two dimensional ferromagnetic case, we assume the magnon states and multi-magnon composite states to be stable in the vicinity of zero temperature. This makes our analytical calculations simpler. Such magnon modes and their interactions are described briefly in the Appendix A. In the following we will make use of the various properties of these magnon states in our novel scheme for the construction of quantum spin vortices and anti-vortices in the flattened meron configuration.

## 2.1 Construction of a quantum spin vortex and anti-vortex

We now start by defining a quantum spin vortex (anti-vortex) on a square lattice [36, 39, 40]. A charge 1 vortex (anti-vortex) is defined on a square plaquette as a spin configuration in which the spin direction (horizontal and vertical spins as defined below) rotates through an angle $+2\pi(-2\pi)$ for a closed walk in an anti-clockwise (clockwise) direction around the plaquette. The vorticity of such a vortex is $+1(-1)$ (see Fig.1 (a) and 1(b)). For our specific model with $\lambda \to 0$ (see eqn. (1)) the in-plane components of spin operators constitute a vortex (anti-vortex). It is worthwhile to mention that this situation corresponding to $\lambda \to 0$ is very different from the case of $\lambda = 0$ corresponding to pure XY model. It is in this very limit that a vortex may be looked upon as a "flattened Meron" [39, 40, 60].

We first assign coordinates $(i,j)a$ ; $(i+1,j)a$; $(i+1,j+1)a$ and $(i,j+1)a$ to the four vertices of the vortex where "$a$" is the lattice parameter. Then the physical realization (spin profile) of the vortex/anti-vortex is determined by the expectation values for the components of the spin operator $S$ on the vertices of the vortex/anti-vortex. The relevant spin states at the vertices are constructed and explained below. For a vortex/anti-vortex having topological charge 1 (in the units of $2\pi$) or simply 1-vortex/1-anti-vortex, the operator expectation values for the different components of spins ($S$) at the vertices are given in the table below,

| $\langle S^x_{ij} \rangle = \frac{1}{2}$, $\langle S^y_{ij} \rangle = 0$, $\langle S^z_{ij} \rangle = 0$ | $\langle S^x_{i+1\ j} \rangle = 0$, $\langle S^y_{i+1\ j} \rangle = \pm\frac{1}{2}$, $\langle S^z_{i+1\ j} \rangle = 0$, +sign for vortex and –sign for anti-vortex |
|---|---|
| $\langle S^x_{i+1\ j+1} \rangle = -\frac{1}{2}$, $\langle S^y_{i+1\ j+1} \rangle = 0$, $\langle S^z_{i+1\ j+1} \rangle = 0$ | $\langle S^x_{i\ j+1} \rangle = 0$, $\langle S^y_{i\ j+1} \rangle = \pm\frac{1}{2}$, $\langle S^z_{i\ j+1} \rangle = 0$, -sign for vortex and +sign for anti-vortex |

**Table: 1**

Let us first construct the quantum state representing a vortex having topological charge '1'. The arrows, representing the spin directions on the four vertices (see Fig. 1(a)), signify that the spin states at the four vertices are such that the expectation values for $S^x, S^y$ and $S^z$ take the values as given in Table:1. The horizontal arrow $|\Rightarrow\rangle$ on the $(i,j)th$ site represents a spin state which is the eigenstate of $S^x_{ij}$ with eigenvalue $+\frac{1}{2}$ and the vertical arrow $|\Uparrow\rangle$ on the $(i+1,j)th$ site represents a spin state which is the eigenstate of $S^y_{ij}$ with eigenvalue $+\frac{1}{2}$. Similarly, the spin state $|\Leftarrow\rangle$ at $(i+1,j+1)th$ site and $|\Downarrow\rangle$ at $(i,j+1)th$ site are the eigenstates of $S^x$ and $S^y$ respectively with the eigenvalue $-\frac{1}{2}$. The spin state corresponding to $|\Rightarrow\rangle$ can be written as a linear combination of the two eigen-states of $S^z$, viz. $|\uparrow\rangle$ and $|\downarrow\rangle$. Then at the $(i,j)'th$ site, the spin state is given by $(a_{ij}|\uparrow\rangle + b_{ij}|\downarrow\rangle)$. The value of $a_{ij}$ and $b_{ij}$ can be determined by using the expectation values for $S^x_{ij}, S^y_{ij}$ and $S^z_{ij}$ and the condition that the eigenvalue of $S^x_{ij}$ is $+\frac{1}{2}$ in the state $(a_{ij}|\uparrow\rangle + b_{ij}|\downarrow\rangle)$. Similarly, for the rest of the vertices corresponding to the vortex the spin states are taken to be of the form $(a|\uparrow\rangle + b|\downarrow\rangle)$.

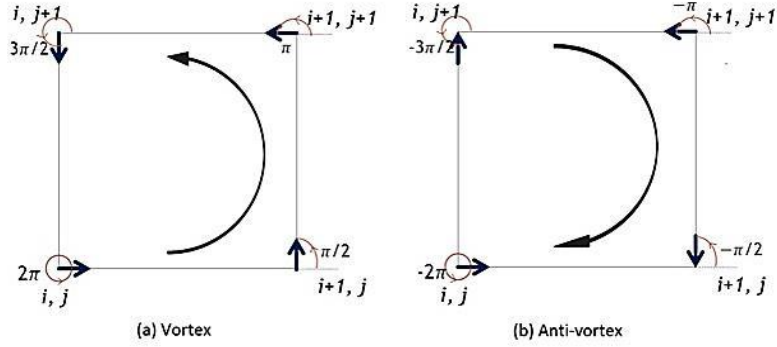

**Figure 1: (a) quantum spin vortex of charge 1, (b) quantum spin anti-vortex of charge 1**

The coefficients $'a'$ and $'b'$ are determined from the expectation values for $S^x$, $S^y$ and $S^z$ and the eigenvalue conditions for the respective vertices as mentioned above. The coefficients $'a'$ and $'b'$ for the four vertices turns out to be,

$$a_{ij} = b_{ij} = \frac{1}{\sqrt{2}} e^{i\theta_{ij}}; \quad a_{i+1\,j} = \frac{1}{\sqrt{2}} e^{i\theta_{i+1\,j}}, b_{i+1\,j} = \frac{i}{\sqrt{2}} e^{i\theta_{i+1\,j}} \tag{5}$$

$$a_{i+1\,j+1} = b_{i+1\,j+1} = \frac{1}{\sqrt{2}} e^{i(\theta_{ij}+\pi)}; \quad a_{i\,j+1} = \frac{1}{\sqrt{2}} e^{i(\theta_{i+1\,j}+\pi)}, b_{i\,j+1} = \frac{i}{\sqrt{2}} e^{i(\theta_{i+1\,j}+\pi)}.$$

Here $\theta_{ij}$ and $\theta_{i+1\,j}$ are arbitrary phase factors and for the diagonally opposite vertices, the coefficients have $\pi$ phase difference. The coefficients given in (5) take care of the proper normalization of the spin states at each vertex. Therefore, the normalized charge-1 vortex state in the background of the original ground state can be defined on a plaquette $((i,j); (i+1,j); (i+1,j+1); (i,j+1))$ as [2],

$$|1V\rangle = |\downarrow\rangle_{11} \otimes |\downarrow\rangle_{12} \otimes |\downarrow\rangle_{13} \otimes \cdots \otimes (a_{ij}|\uparrow\rangle_{ij} + b_{ij}|\downarrow\rangle_{ij}) \otimes (a_{i+1\,j}|\uparrow\rangle_{i+1\,j} + b_{i+1\,j}|\downarrow\rangle_{i+1\,j}) \tag{6}$$
$$\otimes (a_{i+1\,j+1}|\uparrow\rangle_{i+1\,j+1} + b_{i+1\,j+1}|\downarrow\rangle_{i+1\,j+1}) \otimes (a_{i\,j+1}|\uparrow\rangle_{i\,j+1} + b_{i\,j+1}|\downarrow\rangle_{i\,j+1}) \otimes \cdots \cdot |\downarrow\rangle_{NN}.$$

The vorticity operator is defined on the square plaquette as [36],

$$V_{op} = S^x_{ij}S^y_{i+1\,j} - S^y_{i+1\,j}S^x_{i+1\,j+1} + S^x_{i+1\,j+1}S^y_{i\,j+1} - S^y_{i\,j+1}S^x_{ij}. \tag{7}$$

When we operate the vorticity operator on the above 1-vortex state the eigenvalue comes out to be $+1$ as expected.

---

[2] It may be remarked here that in our present formalism we are implicitly assuming a static vortex configuration. This is in conformity with the frozen vortex anti-vortex scenario proposed below $T_{BKT}$ [19, 20].

Similarly we can construct an anti-vortex with charge 1. For an anti-vortex coefficients at the four vertices come out to be,

$$a_{ij} = b_{ij} = \frac{1}{\sqrt{2}} e^{i\theta_{ij}}; \quad a_{i+1\,j} = \frac{1}{\sqrt{2}} e^{i\theta_{i+1\,j}}, b_{i+1\,j} = -\frac{i}{\sqrt{2}} e^{i\theta_{i+1\,j}} \tag{8}$$

$$a_{i+1\,j+1} = b_{i+1\,j+1} = \frac{1}{\sqrt{2}} e^{i(\theta_{ij}+\pi)}; \quad a_{i\,j+1} = \frac{1}{\sqrt{2}} e^{i(\theta_{i+1\,j}+\pi)}, b_{i\,j+1} = -\frac{i}{\sqrt{2}} e^{i(\theta_{i+1\,j}+\pi)}.$$

The explicit structure of the 1-anti-vortex state ($|1AV\rangle$) is same as the state $|1V\rangle$ of (6) with the coefficients $'a'$ and $'b'$ being different from that of the $|1V\rangle$. In this case, the eigenvalue of the vorticity operator defined in (7) comes out to be $-1$, as expected.

## 2.2 Connection between quantum vortex and magnons

Let us consider the simplest situation where there is a single vortex on a $N \times N$ square lattice. To explore the connection between a vortex state and the magnon states, we rewrite the one vortex state in a suitable way. For each vertex of the vortex the spin state $|\uparrow\rangle_{ij}$ can be written as $S_{ij}^+ |\downarrow\rangle_{ij}$, where $S_{ij}^+ = S_{ij}^x + i\, S_{ij}^y$ is the spin raising operator. The state $|1V\rangle$ in eqn. (6) can be rewritten as,

$$\begin{aligned}
|1V\rangle = &\; a_{ij} a_{i+1\,j} a_{i+1\,j+1} a_{i\,j+1} S_{ij}^+ S_{i+1\,j}^+ S_{i+1\,j+1}^+ S_{i\,j+1}^+ |0\rangle \\
&+ \big(b_{ij}\, a_{i+1\,j}\, a_{i+1\,j+1}\, a_{i\,j+1} S_{i+1\,j}^+ S_{i+1\,j+1}^+ S_{i\,j+1}^+ |0\rangle + \cdots \\
&\quad + a_{ij}\, a_{i+1\,j}\, a_{i+1\,j+1}\, b_{i\,j+1} S_{ij}^+ S_{i+1\,j}^+ S_{i+1\,j+1}^+ |0\rangle\big) \\
&+ \big(a_{ij}\, a_{i+1\,j}\, b_{i+1\,j+1}\, b_{i\,j+1} S_{ij}^+ S_{i+1\,j}^+ |0\rangle + \cdots \\
&\quad + a_{ij}\, b_{i+1\,j}\, b_{i+1\,j+1}\, a_{i\,j+1} S_{ij}^+ S_{i\,j+1}^+ |0\rangle\big) \\
&+ \big(a_{ij}\, b_{i+1\,j}\, b_{i+1\,j+1}\, b_{i\,j+1} S_{ij}^+ |0\rangle + b_{ij}\, a_{i+1\,j}\, b_{i+1\,j+1}\, b_{i\,j+1} S_{i+1\,j}^+ |0\rangle \\
&\quad + b_{ij}\, b_{i+1\,j}\, a_{i+1\,j+1}\, b_{i\,j+1} S_{i+1\,j+1}^+ |0\rangle + b_{ij}\, b_{i+1\,j}\, b_{i+1\,j+1}\, a_{i\,j+1} S_{i\,j+1}^+ |0\rangle\big) \\
&+ b_{ij} b_{i+1\,j} b_{i+1\,j+1} b_{i\,j+1} |0\rangle,
\end{aligned} \tag{9}$$

where $a$ and $b's$ are given by eqns. (5). The above equation (9) explicitly shows that the charge 1 quantum vortex state is a linear superposition of one 4-spin deviation state, four 3-spin deviation states, six 2-spin deviation states, four 1-spin deviation states and the ground state. Using the definitions of magnon states, (see Appendix A) the charge 1 vortex state can now be expressed in terms of magnon states as,

$$\begin{aligned}
|1V\rangle = &\; A \sum_{k_1,k_2,k_3,k_4} f_{k_1}^{i,j} f_{k_2}^{i+1,j} f_{k_3}^{i+1,j+1} f_{k_4}^{i,j+1} |k_1 k_2 k_3 k_4\rangle + \sum_{k_1,k_2,k_3} \big(B_1\, f_{k_1}^{i,j} f_{k_2}^{i+1,j} f_{k_3}^{i+1,j+1} \\
&+ B_2 f_{k_1}^{i,j+1} f_{k_2}^{i+1,j} f_{k_3}^{i+1,j+1} + B_3 f_{k_1}^{i,j} f_{k_2}^{i,j+1} f_{k_3}^{i+1,j+1} + B_4 f_{k_1}^{i,j} f_{k_2}^{i,j+1} f_{k_3}^{i+1,j}\big)|k_1 k_2 k_3\rangle \\
&+ \sum_{k_1,k_2} \big(C_1 f_{k_1}^{i,j} f_{k_2}^{i+1,j} + C_2 f_{k_2}^{i+1,j} f_{k_3}^{i+1,j+1} + C_3 f_{k_3}^{i+1,j+1} f_{k_4}^{i,j+1} + C_4 f_{k_1}^{ij} f_{k_4}^{i,j+1} \\
&+ C_5 f_{k_2}^{i+1,j} f_{k_4}^{i,j+1} + C_6 f_{k_1}^{ij} f_{k_3}^{i+1,j+1}\big)|k_1 k_2\rangle + \sum_{k_1} \big(D_1 f_{k_1}^{i,j} + D_2 f_{k_1}^{i+1,j} \\
&+ D_3 f_{k_1}^{i+1,j+1} + D_4 f_{k_1}^{i,j+1}\big)|k_1\rangle + E\,|0\rangle.
\end{aligned} \tag{10}$$

The coefficients $A, B_1, B_2 \cdots$ and $E$ signify the weightage of the different spin deviation states in the composition of charge 1 vortex state and they are given by, $A = a_{ij} a_{i+1\,j} a_{i+1\,j+1} a_{i\,j+1}$, $B_1 = a_{ij}\ a_{i+1\,j}\ a_{i+1\,j+1}\ b_{i\,j+1}, \cdots, E = b_{ij}\ b_{i+1\,j}\ b_{i+1\,j+1}\ b_{i\,j+1}$, where "$a$" and "$b$" are given in eqn. (5)

In case of charge 1 anti-vortex the form of the state $|1AV\rangle$ will be same as the state $|1V\rangle$ except that the values for the coefficients $A = a_{ij} a_{i+1\,j} a_{i+1\,j+1} a_{i\,j+1}, B_1 = a_{ij}\ a_{i+1\,j}\ a_{i+1\,j+1}\ b_{i\,j+1}, \cdots, E = b_{ij}\ b_{i+1\,j}\ b_{i+1\,j+1}\ b_{i\,j+1}$ are different from those for $|1V\rangle$. Their values are determined by the values of $a$ and $b$, as given in eq. (8).

Thus eqn. (10) signifies the fact that the quantum state representing a 1-vortex (1- anti-vortex) is a combination of linear superpositions of 4-magnon composites, 3-magnon composites, 2-magnon composites, 1-magnon states and the ground state.

## 3 Calculations and results regarding stability of the vortex state

In a realistic situation a macroscopic number of vortex anti-vortex pairs needs to be considered. However description of these in terms of multi-magnon composite states will be quite challenging and tough. Therefore as a first step we handle here two special cases, viz., vortex (anti-vortex) in an infinitely dilute limit and also in the finite density limit, as explained below.

Let us now investigate the quantum mechanical stability of the charge 1 vortex state $|1V\rangle$. Operating the Hamiltonian $\mathcal{H}$ (as given by eqn. (1)) on the 1- vortex state $|1V\rangle$ (see eqn. (9)), it can easily be shown that $|1V\rangle$ is not an exact eigenstate for the Hamiltonian $\mathcal{H}$ (see eqn. (11)). Therefore, the natural question arises that how stable the state $|1V\rangle$ is for a system, which is governed by the Hamiltonian $\mathcal{H}$.

For the stability analysis two cases shall be treated separately. In the first one we shall consider the presence of only one charge 1 vortex in the $N \times N$ square lattice. This is the extreme dilute limit where the vortex density is vanishingly small. In the second case we shall consider a finite density of charge 1 vortices to be present in the $N \times N$ square lattice.

### 3.1 Single charge 1 vortex

Let us first consider a single 1-vortex in a $N \times N$ square lattice which is the "extreme dilute limit" of the vortex density. The quantum mechanical state $|1V\rangle$ describing such a situation is given by eqn. (9). Operating the Hamiltonian $\mathcal{H}$ given by eqn. (1), on the state $|1V\rangle$ we get,

$$\mathcal{H}|1V\rangle = (\mathcal{E}_0 + 2\lambda J\hbar^2)|1V\rangle + |\phi_{resi}\rangle \tag{11}$$

(see Appendix B). The right hand side of the above equation clearly shows departure of the vortex state from being an eigenstate of $\mathcal{H}$.

Let us note that the residual state denoted by $|\phi_{resi}\rangle$ is not a linear superposition of multi-magnon states (eqn. (B.1), Appendix B) unlike the state $|1V\rangle$ (eqn. (10)). In fact $|\phi_{resi}\rangle$ (see eqn. (B.2), Appendix B) contains terms which generate higher order inter-multi-magnonic correlations.

The expectation of the Hamiltonian in the state $|1V\rangle$ is evaluated from eqn. (9) and is given by:

$$\langle 1V|\mathcal{H}|1V\rangle = (\mathcal{E}_0 + 3\lambda J\hbar^2) = E_0, \tag{12}$$

where $E_0 \equiv (\mathcal{E}_0 + 3\lambda J\hbar^2)$ and the quantity $3\lambda J\hbar^2$ signifies the energy required to excite one 1-vortex from the ground state, the ground state energy being $\mathcal{E}_0$ as given in (4). The eqn. (11) can be rewritten as,

$$\mathcal{H}|1V\rangle = E_0|1V\rangle + (|\phi_{resi}\rangle - \lambda J\hbar^2|1V\rangle) = E_0|1V\rangle + |\psi_{resi}\rangle, \quad (13)$$

where $|\psi_{resi}\rangle \equiv (|\phi_{resi}\rangle - \lambda J\hbar^2|1V\rangle)$ is again not a linear superposition of multi-magnon states as explained above. Making use of eqns. (11) - (13), it is clear that $\langle 1V|\psi_{resi}\rangle = 0$.

Now operating the Hamiltonian $\mathcal{H}$ successively twice on $|1V\rangle$ the expectation value of $\mathcal{H}^2$ in the state $|1V\rangle$ turns out to be,

$$\langle 1V|\mathcal{H}^2|1V\rangle = (\mathcal{E}_0 + 2\lambda J\hbar^2)^2 + (J\hbar^2)^2 \left(2 + \frac{3}{4}\lambda^2\right). \quad (14)$$

The quantum mechanical stability of the state $|1V\rangle$ is now verified by operating the time evolution operator $[\exp\left(-\frac{i}{\hbar}\mathcal{H}t\right)]$ on the state $|1V\rangle$. Since the state $|1V\rangle$ is not an eigenstate of $\mathcal{H}$ let us take the expectation value of the time evolution operator in $|1V\rangle$, to study what fraction of the original one quantum vortex state is retained during the time evolution of the system. Hence,

$$\langle 1V(0)|1V(t)\rangle = \langle 1V|\exp(-\frac{i}{\hbar}\mathcal{H}t)|1V\rangle = 1 - \frac{it}{\hbar}\langle 1V|\mathcal{H}|1V\rangle + \left(\frac{i}{\hbar}\right)^2\frac{t^2}{2!}\langle 1V|\mathcal{H}^2|1V\rangle + \cdots \quad (15)$$

where $|1V(0)\rangle$ is the initial state and $|1V(t)\rangle$ is the final state (i.e. the after time evolution for a duration of time t). On the right hand side of the above expression we retain terms up-to 2$^{nd}$ order in time explicitly and then eqn. (15) becomes,

$$\langle 1V(0)|1V(t)\rangle = \left(1 - \frac{i}{\hbar}E_0 t + \left(\frac{i}{\hbar}\right)^2\frac{1}{2!}E_0^2 t^2\right) - \frac{1}{\hbar^2}(J\hbar^2)^2\left(1 + \frac{3}{8}\lambda^2\right)t^2 + \mathcal{O}(t^3). \quad (16)$$

It is clear from eqn. (16) that the first three terms correspond to the series expansion of $[\exp(-\frac{i}{\hbar}E_0 t)]$ up-to 2$^{nd}$ order in time. The next one represents the deviation in the sense (of a damping) that in absence of this term, the expectation value of the time evolution operator describes a stationary state exhibiting phase oscillation with frequency $\omega_0 = \frac{E_0}{\hbar}$ and therefore the state $|1V\rangle$ behaves like an approximate eigenstate of the Hamiltonian $\mathcal{H}$ with energy $E_0$. On the other hand the inverse time scale $\Gamma_d$ corresponding to the damping term arising from inter-multi-magnonic correlations as explained above, is given by,

$$\Gamma_d = J\hbar\sqrt{\left(1 + \frac{3}{8}\lambda^2\right)} \quad (17)$$

which essentially indicates the decay rate of the coherent phase oscillation. Hence up-to 2$^{nd}$ order in time, the quantity of interest, viz. the ratio of the decay rate and the phase coherent oscillation frequency comes out to be,

$$\frac{\Gamma_d}{\omega_0} = \frac{1}{\left(\frac{\mathcal{N}}{2} - 3\right)}\sqrt{\left(\frac{1}{\lambda^2} + \frac{3}{8}\right)} \quad (18)$$

In the above ratio the term under the squared root becomes approximately $1/\lambda$ for a very small but fixed value of the anisotropy parameter $\lambda$. Hence eqn. (18) becomes,

$$\frac{\Gamma_d}{\omega_0} \approx \frac{1}{\left(\frac{N}{2} - 3\right)\lambda} \tag{19}$$

The time duration of the evolution is assumed to be much shorter than the natural time scale $t_{nat} = \frac{2\hbar}{\sqrt{3}(J\hbar^2)}$ (for S=1/2) for the system so that a truncation at $2^{nd}$ order in time can be considered safe, where the quantity $J\hbar^2$ has the dimension of energy. At the first place, such an approximation physically means that the multi-magnon composites fuses to form such a vortex state of true quantum nature in a time scale which is much shorter than the natural time scale of the system. Furthermore, the ratio of the evolution time and $t_{nat}$ is assumed to be much smaller than that of the time scale of decay and $t_{nat}$. It is clear from eqn. (19) that $\Gamma_d$ becomes very small compared to $\omega_0$ when the lattice size is very large and ensures the fulfillment of the above conditions. In this case the deviation representing phase incoherence remains ineffective and the state $|1V\rangle$ remains a stable state for the Hamiltonian $\mathcal{H}$. It is worthwhile to mention that in the confined phase below $T_{BKT}$ (and above $T_c$) the form of the dynamical structure function for an ideal vortex gas is a pure delta function $\delta(\omega)$ [35]. However, just above $T_c$ (= 0) taking into consideration the dynamics of all the magnon modes and the multi-magnon composites which are present in a fragile manner, the central peak of the dynamical structure function acquires a finite width. This width expected to be of the order of magnitude of the decay rate $\Gamma_d$.

Let us now estimate the threshold size of the system beyond which the 1-vortex state remains stable. Taking a typical value of $\lambda \approx 10^{-4}$ (since we are considering extreme anisotropy limit, i.e., flattened meron configuration) we have from eqn. (19), $\frac{\Gamma_d}{\omega_0} \approx \frac{2}{N^2\lambda}$. This leads to the above threshold system size to be of the order of $141 \times 141$. Considering a typical value of 3Å for the lattice spacing, the length scale of the system is of the order of $10^{-5}$ cm which falls into the mesoscopic length scale.

## 3.2 Finite density of charge 1 vortices

Since the Hamiltonian contains only the nearest neighbor interactions, the state $\mathcal{H}|1V\rangle$ will produce spin deviations only on the nearest neighbor sites of the vertices of the vortex. Thus to construct a finite density of charge 1 vortices in an infinite lattice, we employ the periodic boundary condition (PBC) for simplicity on a closed (torus) $4 \times 4$ cell, which is of minimum allowed size. By periodically repeating these cells we can fill up the entire $N \times N$ square lattice with a maximum of $n = \frac{N^2}{16}$ (where $N$ is an integral multiple of 4) charge 1 vortices, without having interactions between them, as can be seen in Fig. 2. This is the other limit as opposed to the extreme dilute case studied in Sec. 3.1. The periodicity is therefore, given by the following equations involving spin operators (see Table: 1 under Sec. 3.1),

$$\langle S^\alpha_{i,j}\rangle = \langle S^\alpha_{i+4,j}\rangle = \langle S^\alpha_{i-4,j}\rangle = \langle S^\alpha_{i,j-4}\rangle = \langle S^\alpha_{i,j+4}\rangle, \text{ where } \alpha = x, y, z \tag{20}$$

for all $i$ and $j$ on the lattice, where $\alpha \equiv x, y, z$.

Under these conditions the magnon modes defined in each cell with the PBC will be repeated in the adjacent cell in a periodic manner. Therefore, the composite quantum state corresponding to $'n'$ number of such 1-vortices can be written as,

$$|n1V\rangle = \cdots \otimes |C_{i-4,j}\rangle \otimes |C_{i,j-4}\rangle \otimes |C_{i,j}\rangle \otimes |C_{i+4,j}\rangle \otimes |C_{i,j+4}\rangle \otimes \cdots \tag{21}$$

where the quantum state corresponding to each cell is denoted as $|C_{i,j}\rangle$ which is of the same form as given in eqn. (9) with only exception being the fact that now the number of lattice points is 16 and only four lattice points correspond to the four vertices of the vortex (see Fig.2). Operating the Hamiltonian $\mathcal{H}$ on the state $|n1V\rangle$ and making use of eqns. (13) and (21) we get,

$$\mathcal{H}|n1V\rangle = n\widetilde{E_0}|n1V\rangle \qquad (22)$$
$$+\{|\psi_{resi}^1\rangle \otimes |C_2\rangle \otimes |C_3\rangle \otimes \ldots \otimes |C_n\rangle + |C_1\rangle \otimes |\psi_{resi}^2\rangle \otimes |C_3\rangle \otimes \ldots \otimes |C_n\rangle + \cdots$$
$$+ |C_1\rangle \otimes |C_2\rangle \otimes |C_3\rangle \otimes \ldots \otimes |\psi_{resi}^n\rangle\},$$

Where $\widetilde{E_0} = -5\lambda J\hbar^2$ and the residual state $|\psi_{resi}^i\rangle$ is the deviation of the vortex state from being an eigenstate of $\mathcal{H}$ within the $(ij)$-th cell.

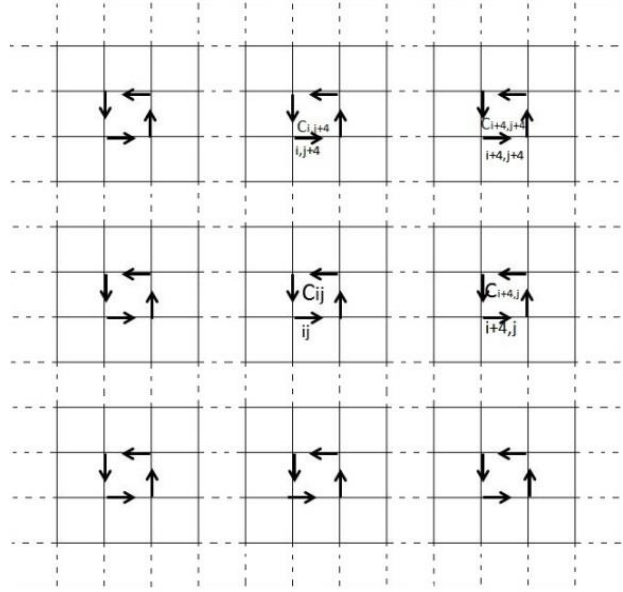

**Figure 2: finite number vortices of charge 1 in an N × N lattice.**

The expression for $\widetilde{E_0}$ corresponding to each $(i,j)$-th cell stands for $E_0$ (see eqn. (20)) with N=4. Also we make use of eqn. (21) with N=4 in deducing eqn. (22). Further we have used the notation $|C_r\rangle$ in place of $|C_{ij}\rangle$ and $|\psi_{resi}^r\rangle$ in place of $|\psi_{resi}^{ij}\rangle$ for convenience.

The expectation value of $\mathcal{H}$ in the state $|n1V\rangle$ is given by,

$$\langle n1V|\mathcal{H}|n1V\rangle = n\widetilde{E_0} \qquad (23)$$

making use of the fact that for each cell $\langle C_r|\psi_{resi}^r\rangle = 0$. This residual state, defined within one cell, is again a non-linear superposition of multi magnon states as explained in Sec. 3.1. Proceeding along the same lines as in Sec. 3.1 we now have,

$$\langle n1V|\mathcal{H}^2|n1V\rangle \approx n^2\widetilde{E_0}^2 + n(J\hbar^2)^2\left(2 + \frac{3}{4}\lambda^2\right) \qquad (24)$$

To check the quantum mechanical stability of the state $|n1V\rangle$ under time evolution we follow the same procedure as adopted in Sec 3.1. Then the overlap between the initial state and the final state (i.e. the expectation value of the time evolution operator $[\exp(-\frac{i}{\hbar}\mathcal{H}t)]$ in the state $|n1V\rangle$ ) comes out to be,

$$\langle n1V(0)|n1V(t)\rangle = \langle n1V\left|\exp\left(-\frac{i}{\hbar}\mathcal{H}t\right)\right|n1V\rangle = \left(1 - \frac{i}{\hbar}nE_0 t + \left(\frac{i}{\hbar}\right)^2 \frac{1}{2!}n^2 E_0^2 t^2\right) \quad (25)$$

$$-\frac{1}{\hbar^2}n(J\hbar^2)^2\left(1+\frac{3}{8}\lambda^2\right)t^2 + \mathcal{O}(t^3)$$

where the exponential series has again been expanded up to 2$^{nd}$ order in time and the justification for such an expansion remains the same as that of Sec 3.1. The ratio of the decay rate $\Gamma_d^{(n)}$ corresponding to the deviation term (the superscript n represents the fact that we are considering finite density of charge 1 vortices) and the frequency $\omega_0^{(n)}$ corresponding to the phase oscillation is given by,

$$\frac{\Gamma_d^{(n)}}{\omega_0^{(n)}} = \frac{4}{5}\sqrt{\frac{1}{N^2\lambda^2}\left(1+\frac{3}{8}\lambda^2\right)} \quad (34)$$

In order for the phase coherent mode to physically survive, it follows from the above equation that the necessary condition is:

$$\Gamma_d^{(n)} < \omega_0^{(n)},$$

which in turn implies,

$$N\lambda > \frac{4}{5}, \text{ for } N > N_c \quad (27)$$

Note that the term $\sqrt{\left(1+\frac{3}{8}\lambda^2\right)}$ in the eqn. (34) above is nearly equal to 1, since our starting model itself is strongly XY anisotropic i.e., $\lambda$ is very small.

The above condition (27) signifies that there must be a critical system size $\mathcal{N}_c = N_c^2$ for a given value of $\lambda$ for the stability of the $n$ 1-vortex state. As before we take $\lambda \approx 10^{-4}$ and in this case the threshold system size comes out to be of the order of $8000 \times 8000$. Taking a typical value of 3Å for the lattice spacing as before, the length scale of the system becomes of the order of $10^{-4}$ cm, which again is in the mesoscopic regime.

Experimental studies of spin dynamics would be quite helpful in verifying our above prediction regarding threshold size.

## 4 Conclusions and discussions

Our analysis firmly establishes that the interaction between collective excitations originating from a strongly anisotropic quantum Heisenberg ferromagnet on two dimensional lattices, can lead to the formation of topological excitations of vortex or anti-vortex type (in a flattened Meron configuration) which are localized objects. These collective excitations could be single magnon as well as multi-magnon composites.

We find that in the situation of an infinitely dilute limit of vortex density, the corresponding 1-vortex state is quantum mechanically stable when the system size exceeds a threshold value, keeping the magnon modes well defined. Similar conclusion holds also for the case with finite density of vortices. The only difference in contrast to the dilute limit case is that, for finite density the threshold size is much larger. It is expected that the above features would remain intact even quantitatively for anti-vortices as well.

Regarding the collective modes referred above it may be remarked that these modes become fragile at any finite temperature. This is because the ferromagnetic Curie temperature $T_c$ identically vanishes on pure two dimensional lattices. Therefore, magnon like collective excitations become fragile at any finite temperature, in analogy with three-dimensional ferromagnets [45-48, 61-63]. For layered systems the interlayer coupling i.e., the exchange interaction between the spins of two nearby layers make $T_c$ non-vanishing. With a very small interlayer coupling, however, $T_c$ still remains quite low and above this transition temperature the collective excitations again become fragile [45-48, 61-63].

It is now a well-established fact that in the paramagnetic phase of the Heisenberg model on 3D spatial lattices the damped propagating modes exist [45-48]. The temporal dependence of the spin-spin correlation function in this paramagnetic phase is diffusive in nature with an oscillatory component present sometimes. This is manifested through the occurrence of the central peak for dynamical structure function (in the constant **q** scan) [45-48]. It should be emphasized that this central peak is fundamentally different from the central peak that we had talked about in relation to the mobile BKT vortices. The primary reason for the truly diffusive or damped propagating behavior of the dynamical structure function is due to the temporal evolution of the spin-spin correlation function, which is governed by various higher order correlation functions with non-trivial temperature and **q** (wave vector) dependence [45-48].

As we stated before, in the case of XY anisotropic Heisenberg models on 2D spatial lattices, the Curie temperature $T_c = 0$; whereas the BKT transition temperature ($T_{BKT}$) is finite. Besides, in the paramagnetic phase above $T_{BKT}$, the vortices (and anti-vortices) move freely and contribute to the dynamical structure function. This provides one of the important mechanisms behind the occurrence of central peak in the dynamical structure function [19-24]. According to our picture the static vortices (anti-vortices) below $T_{BKT}$ are formed from the composite magnon modes which are expected to exist in this temperature regime in a highly fragile manner. Since all the damped propagating composite multi-magnons and the single magnon modes superpose in a complicated manner to form the vortices as shown in this communication, the dynamics of mobile vortices gets very complicated in the regime $T > T_{BKT}$. Our investigation reveals that such a non-trivial combination of all the damped propagating modes gives rise to localized vortex like topological excitations. The temporal behavior of the dynamical spin-spin correlation is further expected to be governed by the various nonlinear processes entering through the higher order correlation functions, bearing the effects of the fragile multi-magnon composites as well.

Apart from forming vortices, some modes are expected to stay intact with their original damped nature. Their dynamics is again either diffusive or damped propagating and can provide a substantial contribution to the central peak both below and above $T_{BKT}$.

Of course, detailed investigations are needed to look into how the diffusive dynamics of the highly damped and fragile multi-magnon composites and the single magnons contribute to the dynamical structure function corresponding to the two-dimensional XY anisotropic Heisenberg spin systems in the temperature regime below and above $T_{BKT}$.

It is however worthwhile to mention that although our investigations pertain to strongly- XY anisotropic case, the material systems of interest to experimentalists mostly belong to weakly- XY anisotropic category [8-11]. In these experimental systems the topological excitations are of truly meronic/ anti-meronic type rather than "flattened meron/ anti-meron" configurations we have dealt with in our calculations.

Last but not least the method of construction of the quantum state representing a charge-1 vortex/anti-vortex has subsequently been extended to the higher charged vortices/ anti-vortices. We find that the quantum state representing any vortex/anti-vortex can be regarded as generated from the interactions between the various magnon modes and magnon-composites [64].

## Acknowledgement:

One of the authors (SS) acknowledges the financial support through Senior Research Fellowship (09/575 (0089)/2010 EMR–1) provided by Council of Scientific and Industrial Research (CSIR), Govt. of India.

## Appendix A

In this paper we are considering only the two-dimensional systems and in a very small temperature regime above zero temperature and the entire soup of magnon-like fragile modes and the composite magnon modes are expected to be found [45-48]. In this appendix magnon states and the interactions between the magnon modes will be reviewed briefly to develop notations for our convenience.

**One Magnon States**: When a spin-deviation is introduced on a particular site of the lattice it does not remain localized on that site. It rather propagates through the lattice due to the exchange interaction between the nearest neighbor spins and thereby constitutes the "spin wave" [52-56, 65]. The basic unit of the quantized spin waves is the Magnons. The normalized quantum state of one spin-deviation is defined as,

$$|ij\rangle = \frac{1}{\sqrt{2S\hbar}} S_{ij}^+ |0\rangle. \tag{A.1}$$

There are $\mathcal{N}(= N^2)$ such orthogonal and normalized states containing one spin deviation each corresponding to all choices of the lattice points. For spin ½ systems, assuming the translational invariance and the periodic boundary condition the one magnon state is defined as,

$$|\mathbf{k}\rangle = \sum_{i,j} \frac{e^{i\mathbf{k}\cdot R_{ij}}}{\sqrt{\mathcal{N}}} S_{ij}^+ |0\rangle = \sum_{i,j} \left(f_{\mathbf{k}}^{i,j}\right)^* S_{ij}^+ |0\rangle \tag{A.2}$$

where $\mathbf{k}$ is the Bloch wave vector restricted in the first Brillouin zone, describing the propagation of the magnon, $\mathbf{R}_{ij}$ is the position vector of $ij^{th}$ lattice site on the square lattice and $f_{\mathbf{k}}^{ij} = \frac{e^{-i\mathbf{k}\cdot R_{ij}}}{\sqrt{\mathcal{N}}}$. The one magnon states defined above are normalized to unity, i.e., $\langle \mathbf{k}|\mathbf{k}'\rangle = \delta_{\mathbf{k}\mathbf{k}'}$ and $|\mathbf{k}\rangle$ forms a complete set of orthonormal states. The $|\mathbf{k}\rangle$ States are the exact eigen-states of the Hamiltonian $\mathcal{H}$ corresponding to eqn. (1) with the eigenvalue $\mathcal{E}_0 + \hbar\omega(\mathbf{k})$

The one magnon excitation energy $\hbar\omega(\mathbf{k})$, above the ground state, is given by,

$$\omega(\mathbf{k}) = 2\hbar J (\lambda - \gamma_\mathbf{k}), \tag{A.3}$$

where $\gamma_\mathbf{k} = \frac{1}{4}\sum_\delta e^{i\mathbf{k}\cdot\delta}$ and $\boldsymbol{\delta}$ is a vector connecting a typical site to its nearest neighbors. The one spin deviation states can be obtained from eqn. (4) by the inverse transformation [52-56, 65].

In the long range ordered phase below the transition temperature ($T_C$ or $T_N$), as the number of magnon increases with the increasing temperature they are more prone to interact with each other and therefore, the composite magnon modes are very natural to occur [54, 55]. This happens when the spatial lattice is three-dimensional. In the following, we shall restate the well-known definitions of composite magnon states [52-56, 65]. The magnon-magnon interactions shall be treated briefly in the Appendix A.

**Two-magnon states:** The two-magnon states can be defined, in a similar manner as the one magnon state (see eqn. (4)) as follows,

$$|k, k'\rangle = \sum_{i,j;p,q} \frac{e^{i(k \cdot R_{ij} + k' \cdot R_{pq})}}{(\sqrt{\mathcal{N}})^2} S_{ij}^+ S_{pq}^+ |0\rangle = \sum_{i,j;p,q} \left(f_k^{i,j}\right)^* \left(f_{k'}^{p,q}\right)^* S_{ij}^+ S_{pq}^+ |0\rangle \quad (A.4)$$

The 2-spin-deviation states $|ij, pq\rangle$ are related to $|k, k'\rangle$ in the following way,

$$S_{ij}^+ S_{pq}^+ |0\rangle = \sum_{k,k'} f_k^{i,j} f_{k'}^{p,q} |k, k'\rangle, \quad (A.5)$$

These two-magnon states are approximately orthonormal with an error of no more than $\mathcal{O}(\frac{1}{\mathcal{N}})$ which can be seen from the form of the scalar product, viz., $\langle kk'|\lambda\lambda'\rangle = \hbar^4 \delta_{\lambda+\lambda',k+k'}(\delta_{\lambda k} + \delta_{\lambda' k'} - \frac{2}{\mathcal{N}})$. The very choice of the form of the two-magnon state (eqn. (7)) leads to what are called Dyson's "kinematical" and "dynamical" interactions [52-56, 65].

**Higher magnon states:** Using the analogous scheme the 3-magnon composite states are defined as,

$$|k_1, k_2, k_3\rangle = \sum_{i,j;,p,q;r,s} \left(f_{k_1}^{i,j}\right)^* \left(f_{k_2}^{p,q}\right)^* \left(f_{k_3}^{r,s}\right)^* S_{ij}^+ S_{pq}^+ S_{rs}^+ |0\rangle \quad (A.6)$$

The 3-spin-deviations state, $S_{ij}^+ S_{pq}^+ S_{rs}^+ |0\rangle$ is defined as the inverse transformation of the 3-magnon which is similar to the definition of the two spin-deviations in eqn. (8) [66-68].

The quantum state of 4-magnon composites can be defined analogously as,

$$|k_1, k_2, k_3, k_4\rangle = \sum_{\substack{i,j;,p,q;\\r,s;l,m}} \left(f_{k_1}^{i,j}\right)^* \left(f_{k_2}^{p,q}\right)^* \left(f_{k_3}^{r,s}\right)^* \left(f_{k_4}^{l,m}\right)^* S_{ij}^+ S_{pq}^+ S_{rs}^+ S_{lm}^+ |0\rangle \quad (A.7)$$

and 4-spin-deviations state $S_{ij}^+ S_{pq}^+ S_{rs}^+ S_{lm}^+ |0\rangle$ is defined as the inverse transformation of the 4-magnon composite states. The simultaneous spin deviations on the direct lattice are governed by the nearest neighbor interaction between the spins.

The set of two-magnon states defined in (7) has the scalar product $\langle kk'|\lambda\lambda'\rangle = \hbar^4 \delta_{\lambda+\lambda',k+k'}(\delta_{\lambda k} + \delta_{\lambda' k'} - \frac{2}{\mathcal{N}})$ and therefore two distinct state vectors are not orthogonal in general. These two-magnon states are approximately orthonormal with an error of no more than $\mathcal{O}(\frac{1}{\mathcal{N}})$ [54, 55, 65]. The effect of the Hamiltonian $\mathcal{H}$ operating on $|k\,k'\rangle$ is given by,

$$\mathcal{H}|k\,k'\rangle = [\mathcal{E}_0 + \hbar\omega(k) + \hbar\omega(k')]|k\,k'\rangle + 2\lambda J\hbar^2 \sum_{i,j;\,\delta} f^{ij}_{k+k'}(1 - \lambda e^{ik'\cdot\delta})|ij, ij+\delta\rangle$$

$$= [\mathcal{E}_0 + \hbar\omega(k) + \hbar\omega(k')]|k\,k'\rangle + \frac{2J\hbar^2}{\mathcal{N}} \sum_{\widetilde{k},\widetilde{k'};\,\delta} \delta_{k+k',\widetilde{k}+\widetilde{k'}}\, e^{-i\widetilde{k'}\cdot\delta}(1 - \lambda e^{ik'\cdot\delta})|\widetilde{k},\widetilde{k'}\rangle, \quad (A.8)$$

where $f^{ij}_k = \frac{e^{-ik\cdot R_{ij}}}{\mathcal{N}}$ as in (4) (see sec. 2.1). The above equation can be rewritten in a convenient form,

$$\mathcal{H}|k\,k'\rangle = [\mathcal{E}_0 + \hbar\omega(k) + \hbar\omega(k')]|k\,k'\rangle + \frac{1}{\mathcal{N}}[\sum_{\widetilde{k},\widetilde{k'}} g_{2M}(k,k';\widetilde{k},\widetilde{k'})|\widetilde{k},\widetilde{k'}\rangle], \quad (A.9)$$

where $g_{2M} = 2J\hbar^2 \sum_\delta \delta_{k+k',\widetilde{k}+\widetilde{k'}}\, e^{-i\widetilde{k'}\cdot\delta}(1 - \lambda e^{ik'\cdot\delta})$ and $\delta$ is a vector connecting a typical lattice site to its nearest neighbours. The last term in the equation (A.2) represents the deviation of the 2-magnon state $|k\,k'\rangle$ from being an eigen-state of the Hamiltonian $\mathcal{H}$. The 2-magnon energy $\mathcal{E}_{2M}(k,k')$ is defined as $\mathcal{E}_{2M}(k,k') = \frac{\langle kk'|\mathcal{H}|kk'\rangle}{\langle kk'|kk'\rangle}$ and is given by,

$$\mathcal{E}_{2M}(k,k') = \mathcal{E}_0 + \hbar\omega(k) + \hbar\omega(k') + \frac{1}{\mathcal{N}}\delta\mathcal{E}_{2M}(k,k'), \quad (A.10)$$

within an error of $\mathcal{O}(\frac{1}{\mathcal{N}})$ [54, 55, 64]. Where the quantity $\delta\mathcal{E}_{2M}(k,k')$ is given by,

$$\delta\mathcal{E}_{2M}(k,k') = 2J\hbar^2 \sum_{\widetilde{k},\widetilde{k'};\,\delta} \delta_{k+k',\widetilde{k}+\widetilde{k'}}\, e^{-i\widetilde{k'}\cdot\delta}(1 - \lambda e^{ik'\cdot\delta}) \frac{\langle k,k'|\widetilde{k},\widetilde{k'}\rangle}{\langle kk'|kk'\rangle} \quad (A.11)$$

The very choice of the form of the two-magnon state (eqn. (7)) leads to what are called Dyson's "kinematical" and "dynamical" interactions [52-55]. The term $\delta\mathcal{E}_{2M}(k,k')$ is in general a complex quantity whose real part represents the interaction energy between two 1-magnons. The imaginary part is related to the inverse scattering lifetime of a given 1-magnon ($k$) in the presence of a finite, but low, density of other excitations [54, 55, 65]. Calculation of the complex binary interaction term $\delta\mathcal{E}_{2M}(k,k')$ is not necessary for our present purpose.

A straightforward generalization of (A.2) for the 3-magnon states is given by,

$$\mathcal{H}|k_1,k_2,k_3\rangle = [\mathcal{E}_0 + \hbar\omega(k_1) + \hbar\omega(k_2) + \hbar\omega(k_3)]|k_1,k_2,k_3\rangle \quad (A.12)$$
$$+ \frac{1}{\mathcal{N}}[\sum_{\widetilde{k_1},\widetilde{k_2},\widetilde{k_3}} g_{3M}(k_1,k_2,k_3;\widetilde{k_1},\widetilde{k_2},\widetilde{k_3})|\widetilde{k_1},\widetilde{k_2},\widetilde{k_3}\rangle]$$

Energy corresponding to the 3-magnon states is given by,

$$\mathcal{E}_{3M}(k_1,k_2,k_3) = \mathcal{E}_0 + \hbar\omega(k_1) + \hbar\omega(k_2) + \hbar\omega(k_3) + \frac{1}{\mathcal{N}}\delta\mathcal{E}_{3M}(k_1,k_2,k_3), \quad (A.13)$$

within an error of $\mathcal{O}(\frac{1}{\mathcal{N}})$. The term $\delta\mathcal{E}_{3M}(k_1,k_2,k_3)$ is in general complex and represents three-magnon interactions corresponding to three simultaneous spin deviations on the direct lattice.

The effect of the Hamiltonian $\mathcal{H}$ operating on the 4-magnon states (given by (10)) is,

$$\mathcal{H}|k_1, k_2, k_3, k_4\rangle \tag{A.14}$$
$$= [\mathcal{E}_0 + \hbar\omega(k_1) + \hbar\omega(k_2) + \hbar\omega(k_3) + \hbar\omega(k_4)]|k_1, k_2, k_3, k_4\rangle$$
$$+ \frac{1}{\mathcal{N}}[\sum_{\widetilde{k_1},\widetilde{k_2},\widetilde{k_3},\widetilde{k_4}} g_{4M}(k_1, k_2, k_3, k_4; \widetilde{k_1}, \widetilde{k_2}, \widetilde{k_3}, \widetilde{k_4})|\widetilde{k_1}, \widetilde{k_2}, \widetilde{k_3}, \widetilde{k_4}\rangle]$$

Similarly, the energy corresponding to the 4-magnon states are given, within an error of $\mathcal{O}(\frac{1}{\mathcal{N}})$ as,

$$\mathcal{E}_{4M}(k_1, k_2, k_3, k_4) \tag{A.15}$$
$$= \mathcal{E}_0 + \hbar\omega(k_1) + \hbar\omega(k_2) + \hbar\omega(k_3) + \hbar\omega(k_4) + \frac{1}{\mathcal{N}}\delta\mathcal{E}_{4M}(k_1, k_2, k_3, k_4)$$

Here the quantity $\delta\mathcal{E}_{4M}(k_1, k_2, k_3, k_4)$ represents 4-magnon interactions corresponding to four simultaneous spin deviations on the direct lattice.

## Appendix B

Operating the Hamiltonian $\mathcal{H}$ (eqn. (1)) on the quantum state corresponding to charge 1 vortex (as given in eqn. (9)) and using the eqns. (4), (A.3), (A.9), (A.12) and (A.14) we find,

$$\mathcal{H}|1V\rangle = (\mathcal{E}_0 + 2\lambda J\hbar^2)|1V\rangle + |\phi_{resi}\rangle \tag{B.1}$$

where the state $|\phi_{resi}\rangle$ (corresponding to eqns. (11) as well as (B.1)) is given by,

$$|\phi_{resi}\rangle = A\left[\sum_{k_1,k_2,k_3,k_4} f_{k_1}^{i,j} f_{k_2}^{i+1,j} f_{k_3}^{i+1,j+1} f_{k_4}^{i,j+1} \left\{2J\hbar^2(3\lambda - \gamma(k_1) - \gamma(k_2) - \gamma(k_3) - \gamma(k_4))|k_1 k_2 k_3 k_4\rangle\right.\right.$$
$$\left.\left.+ \frac{1}{\mathcal{N}} \sum_{\widetilde{k_1},\widetilde{k_2},\widetilde{k_3},\widetilde{k_4}} g_{4M}(k_1, k_2, k_3, k_4; \widetilde{k_1}, \widetilde{k_2}, \widetilde{k_3}, \widetilde{k_4})|\widetilde{k_1}\widetilde{k_2}\widetilde{k_3}\widetilde{k_4}\rangle\right\}\right] +$$

$$\left[\sum_{k_1,k_2,k_3} \left(B_1\ f_{k_1}^{i,j} f_{k_2}^{i+1,j} f_{k_3}^{i+1,j+1} + B_2 f_{k_1}^{i,j+1} f_{k_2}^{i+1,j} f_{k_3}^{i+1,j+1} + B_3\ f_{k_1}^{i,j} f_{k_2}^{i,j+1} f_{k_3}^{i+1,j+1}\right.\right.$$
$$\left.\left.+ B_4\ f_{k_1}^{i,j} f_{k_2}^{i,j+1} f_{k_3}^{i+1,j}\right)\right) \left\{2J\hbar^2(2\lambda - \gamma(k_1) - \gamma(k_2) - \gamma(k_3))|k_1, k_2, k_3\rangle\right.$$
$$\left.+ \frac{1}{\mathcal{N}} \sum_{\widetilde{k_1},\widetilde{k_2},\widetilde{k_3}} g_{3M}(k_1, k_2, k_3; \widetilde{k_1}, \widetilde{k_2}, \widetilde{k_3})|\widetilde{k_1}\widetilde{k_2}\widetilde{k_3}\rangle\right\}\right] +$$

$$\left[\sum_{k_1,k_2}\left(C_1 f^{i,j}_{k_1} f^{i+1,j}_{k_2} + C_2 f^{i+1,j}_{k_2} f^{i+1,j+1}_{k_3} + C_3 f^{i+1,j+1}_{k_3} f^{i,j+1}_{k_4} + C_4 f^{ij}_{k_1} f^{i,j+1}_{k_4} + C_5 f^{i+1,j}_{k_2} f^{i,j+1}_{k_4}\right.\right.$$

$$\left.\left. + C_6 f^{ij}_{k_1} f^{i+1,j+1}_{k_3}\right)\left\{2J\hbar^2(\lambda - \gamma(k_1) - \gamma(k_2))|k_1 k_2\rangle\right.\right.$$

$$\left.\left. + \frac{1}{\mathcal{N}}\sum_{\widetilde{k_1},\widetilde{k_2}} g_{2M}(k_1, k_2; \widetilde{k_1}, \widetilde{k_2})|\widetilde{k_1}\widetilde{k_2}\rangle\right\}\right] +$$

$$\left[\sum_{k_1}(D_1 f^{i,j}_{k_1} + D_2 f^{i+1,j}_{k_1} + D_3 f^{i+1,j+1}_{k_1} + D_4 f^{i,j+1}_{k_1})\left[-\gamma(k_1)\right]|k_1\rangle\right] + (-E)2\lambda J\hbar^2 |0\rangle. \quad (B.2)$$

The significance of the above residual state $|\phi_{resi}\rangle$ has been explained in Sec. 2.1. In the eqn. (B.1) the first term on the right hand side corresponds to nonlinear superposition of 4-magnon composites where the term $g_{4M}(k_1, k_2, k_3, k_4; \widetilde{k_1}, \widetilde{k_2}, \widetilde{k_3}, \widetilde{k_4})$, being in general a complex function, represents the interactions between the four magnon modes. Similarly, the second and third terms correspond to the nonlinear superposition of 3-magnon and 2-magnon composites respectively. The terms $g_{3M}(k_1, k_2, k_3; \widetilde{k_1}, \widetilde{k_2}, \widetilde{k_3})$ and $g_{2M}(k_1, k_2; \widetilde{k_1}, \widetilde{k_2})$ represent the interactions between three magnon modes (identified by the subscript '3M') and two magnon modes (identified by the subscript '2M') respectively which are in general complex functions. The exact expression for $g_{2M}(k_1, k_2; \widetilde{k_1}, \widetilde{k_2})$ has been given in Appendix: A (see eqn. (A.9)). The fourth term in the eqn. (B.1) represents the contribution from the linear superposition of all the 1-magnon modes to the residual state $|\phi_{resi}\rangle$ and the last term gives the ground state contribution. Hence, altogether the residual states $|\phi_{resi}\rangle$ and $|\psi_{resi}\rangle \equiv |\phi_{resi}\rangle - \lambda J\hbar^2 |1V\rangle$ (see eqn. (13)) are arising from the inter-multi-magnonic correlations as stated in Sec. (2.1). The quantum state $|\psi^i_{resi}\rangle$ corresponding to eqn. (22) is obtained by replacing $\mathcal{N}$ by 16 in the above equation and then using the relation $|\psi^i_{resi}\rangle \equiv |\phi^i_{resi}\rangle - \lambda J\hbar^2 |C_i\rangle$ (see Sec. 2.2).